\documentclass[10pt]{article}
\usepackage[OE]{express}

\begin{document}

\title{Tunable ultra-high-efficiency light absorption of monolayer graphene using critical coupling with guided resonance}

\author{Xiaoyun Jiang, Tao Wang,\authormark{*} Shuyuan Xiao, Xicheng Yan, and Le Cheng}

\address{Wuhan National Laboratory for Optoelectronics, Huazhong University of
Science and Technology, Wuhan 430074, People's Republic of China.}

\email{\authormark{*}wangtao@hust.edu.cn} 



\begin{abstract}
We numerically demonstrate a novel monolayer graphene-based perfect absorption multi-layer photonic structure by the mechanism of critical coupling with guided resonance, in which the absorption of graphene can significantly close to $99\%$ at telecommunication wavelengths. The highly efficient absorption and spectral selectivity can be obtained with designing structural parameters in the near infrared ranges. Compared to previous works, we achieve the complete absorption of single-atomic-layer graphene in the perfect absorber for the first time, which not only opens up new methods of enhancing the light-graphene interaction, but also makes for practical applications in high-performance optoelectronic devices, such as modulators and sensors. 
\end{abstract}



\section{Introduction}
Graphene, a novel two-dimensional material, has attracted particular attention recently due to its exceptional optical and electronic properties\cite{novoselov2004electric}. The ultra-broad spectral response, the ultra-thin atomic layer thickness and the ultra-high carrier mobility of graphene make it an ideal material for optoelectronic devices such as
photodetectors\cite{liu2014graphene,song2013great,zhang2015towards,xiao2016tunable},  biosensors\cite{rodrigo2015mid,grande2014graphene,li2016monolayer,xiao2017strong}, and modulators\cite{liu2011graphene,bao2012graphene,he2015tunable,he2015graphene,zhao2016graphene}.  However, for monolayer graphene, there are two inherent defects that hinder its high-performance on optical devices. First, the absorption of monolayer graphene is only $2.3\%$ in the visible to near-infrared ranges, which limits the quantum efficiency and results in low photoresponsivity\cite{nair2008fine}. Second, monolayer graphene does not display spectral selectivity because of its ultra-wide absorption spectrum range from the ultraviolet to the terahertz. Over the past few years, various photonic technologies have been presented to improve the absorption of the monolayer graphene by enhancing the light-graphene interaction. On the one hand, in the visible and near-infrared, one can place monolayer graphene inside various nano- or micro-cavities to achieve the perfect absorption of graphene but the devices are quite complex\cite{gan2013chip,shiue2013enhanced}. On the other hand, Tamm plasmon polaritons (TPPs) and localized plasmons of metallic nanostructures have been used for light trapping to enhance the absorption of the monolayer graphene at communication wavelengths\cite{lu2015highly,lu2016tunable}. However, the metal attenuation and surface reflection lead to a failure to achieve total absorption in monolayer graphene. Therefore, the perfect absorption of the monolayer-graphene is still rare and in urgent need for graphene functional design, especially in the visible and near infrared bands.

In this work, we theoretically investigate a graphene-based perfect absorption structure by using critical coupling with guided resonance theory, in which the absorption of monolayer graphene can reach almost $99\%$ at telecommunication wavelengths. These results originate from the electric field distributions surrounding monolayer graphene can be significantly enhanced by coupling mode with guided resonance of lossless multilayer dielectric combinations. Compared to the previous devices with a metallic reflector, we choice a dielectric Bragg mirrors with fewer layers as back reflector, because the metal parasitic absorption and its own attenuation reduce the light absorption of graphene\cite{grande2015graphene,guo2016experimental}. In addition, the proposed structure is simple and ultra-high-efficiency light absorption of graphene can be achieved by the mechanism of critical coupling. Meanwhile, the selectivity of the spectrum also can be obtained by adjusting the parameters of the structure.
\begin{figure}[htbp]
\centering
\includegraphics[scale=0.35]{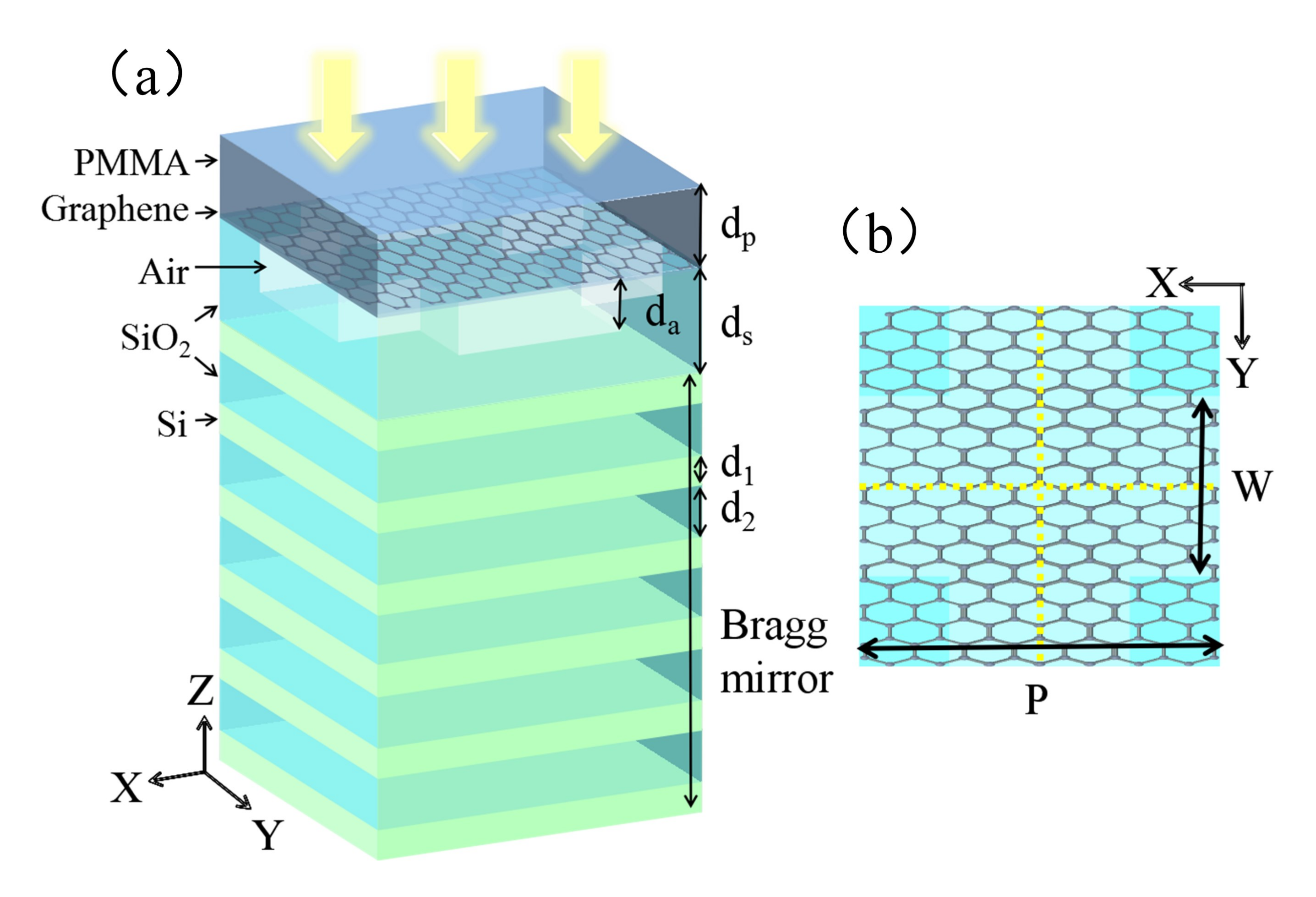}
\caption{\label{fig:1} (a) Schematic  drawing  of the  proposed a monolayer graphene-based perfect absorption structure. (b)  A top view of the designed structure. The yellow cross-shaped dotted lines stand for symmetrical positions. }
\end{figure}

\section{The geometric structure and numerical model}
The schematic image of perfect absorption system with monolayer graphene is shown in Fig.1, a monolayer graphene is sandwiched between a $2$D polynethy $1$-methacrylate (PMMA) layer and a silicon dioxide (SiO$_{2}$) layer with an array of cross-shaped groove air waveguide, and a dielectric Bragg mirror with $5.5$-pair alternately stacked silicon ( Si) and SiO$_{2}$ layers is deposited at the back side of SiO$_{2}$ layer to prevent the transmission of the incident light\cite{barrios2004compact}. Numerical simulations are analyzed by utilizing the finite-difference time-domain (FDTD) method. In the simulation, the monolayer graphene with a thin thickness of $d_G = 1$ nm can be viewed as a conductive surface with a light conductivity of $G_0\approx6.08\times10^{-5}\Omega^{-1}$, which corresponds to free standing graphene absorbs $2.3\%$ of the incident light at the same wavelength\cite{fan2017monolayer}. The refractive indices of PMMA, air, SiO$_{2}$ and Si are taken to be $1.48$, $1$, $1.45$ and $3.48$, respectively. The relative geometrical parameters are labeled on Fig.1.
\begin{figure}[htbp]\label{fig:2}
\centering
\includegraphics[scale=0.3]{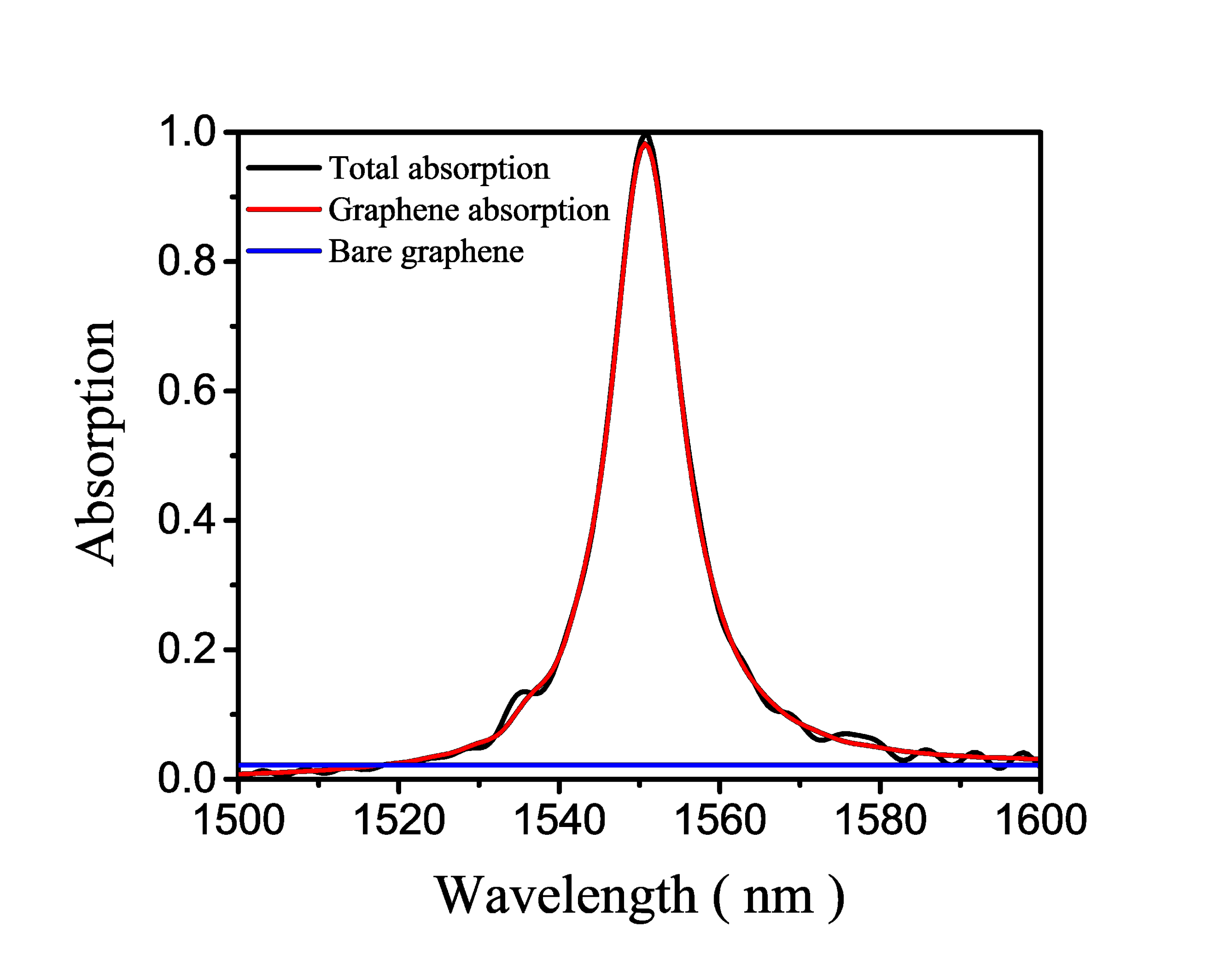}
\caption{ The absorption spectra of the whole designed devices ( black line ) and the monolayer graphene in the structure ( red line ) are compared with bare graphene monolayer standing in air ( blue line ) at the same wavelengths range.}
\end{figure}

\section{Results and discussions}
First of all, in the FDTD simulations, the normal incidence light is supposed to be TM-polarized ( the electric field parallel to the $X$-axis ). Fig. 2 shows that the absorption of the entire structure and monolayer graphene are compared to that of bare graphene in air, when the parameters are assumed as $d_p=440$ nm, $d_s=560$ nm, $d_a=280$ nm, $d_1=100$ nm, $d_2=260$ nm, $w=560$ nm and $P=1250$ nm. We set $N=5.5$ pairs as the dielectric layers in the Bragg mirror in order to make the system more stable and efficient. As depicted in Fig.2, it can be seen that the incident light is almost completely absorbed by the entire device (black line) and monolayer graphene with being inserted in the structure (red line) at resonance wavelength (at communication wavelength of $1550$ nm), compared to the absorption of bare graphene (blue line) standing in air (about $2.3\%$ of the incident light) at the same band. As can be seen from the diagram, the red line and the black line are almost coincident, which indicates that the incident light is totally absorbed by the graphene in the structure.
In order to account for this phenomenon, we will use the coupled mode theory with guided resonance formalism. The coupling mode theory is used to explain the input and output performance of a resonator, which affects coherence directly and indirectly. Since we chose subwavelength structure, only the zero-order mode will propagate, indicating that the incident light will excite a guided resonance at normal incidence, which corresponds to only one absorption peak in Fig.2. We consider a resonator with a single resonance at $\omega_0$ , whose input and output waves of amplitudes are $u$ and $y$, respectively. The external leakage rate of the resonant cavity is $\gamma_e$ , and the intrinsic loss rate of monolayer graphene is $\delta$, the reflectivity coefficient of the system can be calculated by the equation\cite{piper2014total}, 
\begin{equation}\label{eq1}
    \Gamma(\omega)\equiv\frac{y}{u}=\frac{j(\omega-\omega_{0})+\delta-\gamma_{e}}{j(\omega-\omega_{0})+\delta+\gamma_{e}},
\end{equation}
and the absorption can be defined by the equation,
\begin{equation}\label{eq2}
    \begin{split}
     A(\omega) &=1-\mid\Gamma(\omega)\mid^2     \\
               &=\frac{4\delta\gamma_{e}}{(\omega-\omega_{0})^2+(\delta+\gamma_{e})^2}.
    \end{split}
\end{equation}
From eq$1$ and eq$2$, it can be seen that when the system is in the resonance state ($\omega=\omega_0$), and the external leakage rate is equal to the intrinsic loss rate of graphene ($\gamma_e=\delta$), the whole system satisfies the critical coupling condition at which the reflection coefficient vanishes and all incident energies are absorbed. In addition, monolayer graphene has low single-pass and high transmittance at communication wavelengths, making it a minimum disruption underlying the behaviour of the resonator, thus, we can use the guided resonance to obtain the critical coupling of graphene to enhance its absorption rate. In other words, when the system meets the critical coupling condition ($\gamma_e=\delta$) and the guided resonance is excited in the cross-shaped air groove with the incident light at resonance wavelength, the electric field intensity around the monolayer graphene is enhanced by the guided resonance of a cross-shaped groove resonator, which reinforce the graphene-light interaction and boost the absorption of graphene. As shown in Fig.3a and b, when the resonant cavity is excited (on-resonant) and satisfies the critical coupling condition, corresponding to the peak absorption ($1550$ nm) in Fig.2, the electric field intensity distribution at this time is shown in Fig.3a, and the electric field intensity around the graphene is obviously enhanced. In contrast, when the resonant cavity is not excited (off-resonant), the reflection coefficient of the system can be equivalent to $1$, corresponding to the low absorption value ($1600$ nm) in Fig.2, and the electric field intensity distribution is shown in Fig.3b.
\begin{figure}[htbp]\label{fig:3}
\centering
\includegraphics[scale=0.4]{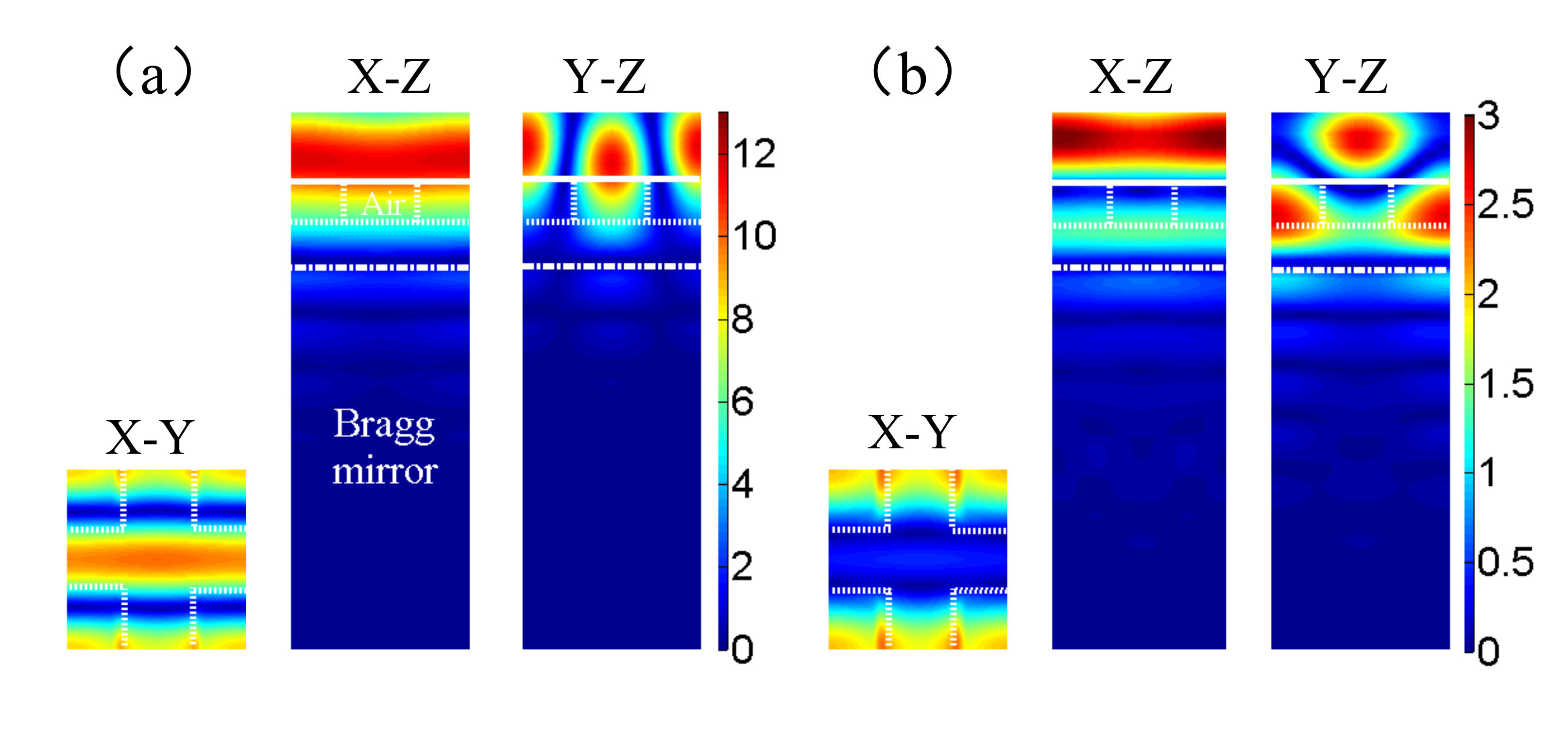}
\caption{ Simulated electric field amplitude distributions of the proposed a graphene-based structure under normal incidence at on-resonant ($1550$ nm) wavelength (a) and off-resonant ($1600$ nm) wavelength (b). The location of the solid lines stand for the vicinity of monolayer graphene and the dotted lines represent the air guide cavity, while under the dashed-dotted lines represent Bragg mirror.}
\end{figure}

Considering that the absorption of monolayer graphene within the band range of our study is largely independent of frequency, and it has a relatively fixed intrinsic loss rate($\delta$), therefore, controlling the external leakage rate ($\gamma_e$) of the structure is the key to realize the perfect absorption of graphene. Here , we investigate the relationship between the external leakage rate ($\gamma_e$) and various related structural parameters, and effect of changing parameters on graphene absorption. As shown in Fig.4a and b, when the width ($w$) and depth ($d_a$) of the cross-shaped groove air resonator are adjusted, the corresponding absorption spectrum of the monolayer graphene undergoes a significant blue shift. Meanwhile, the peaks magnitude of the absorption are also altered. The main reason is that the external leakage rate ($\gamma_e$) of the resonator increases continuously, and the system experience three states, namely, undercoupling, critical coupling and overcoupling. Monolayer-graphene perfect absorption can only occur in the critical coupling state, corresponding to the spectral lines of $w=560$ nm and $d_a=280$ nm in the figure. At the same time, we also consider the effect of SiO$_{2}$ ($d_s$) thickness in the system as shown in Fig.4c. We can find that the thickness of SiO$_{2}$ has a small effect on the absorption of graphene and resonance wavelength compared to the previous two parameters ($w$ and $d_a$), this is because the external leakage rate ($\gamma_e$) is not sensitive to its minor changes.
\begin{figure}[htbp]\label{fig:4}
\centering
\includegraphics[scale=0.32]{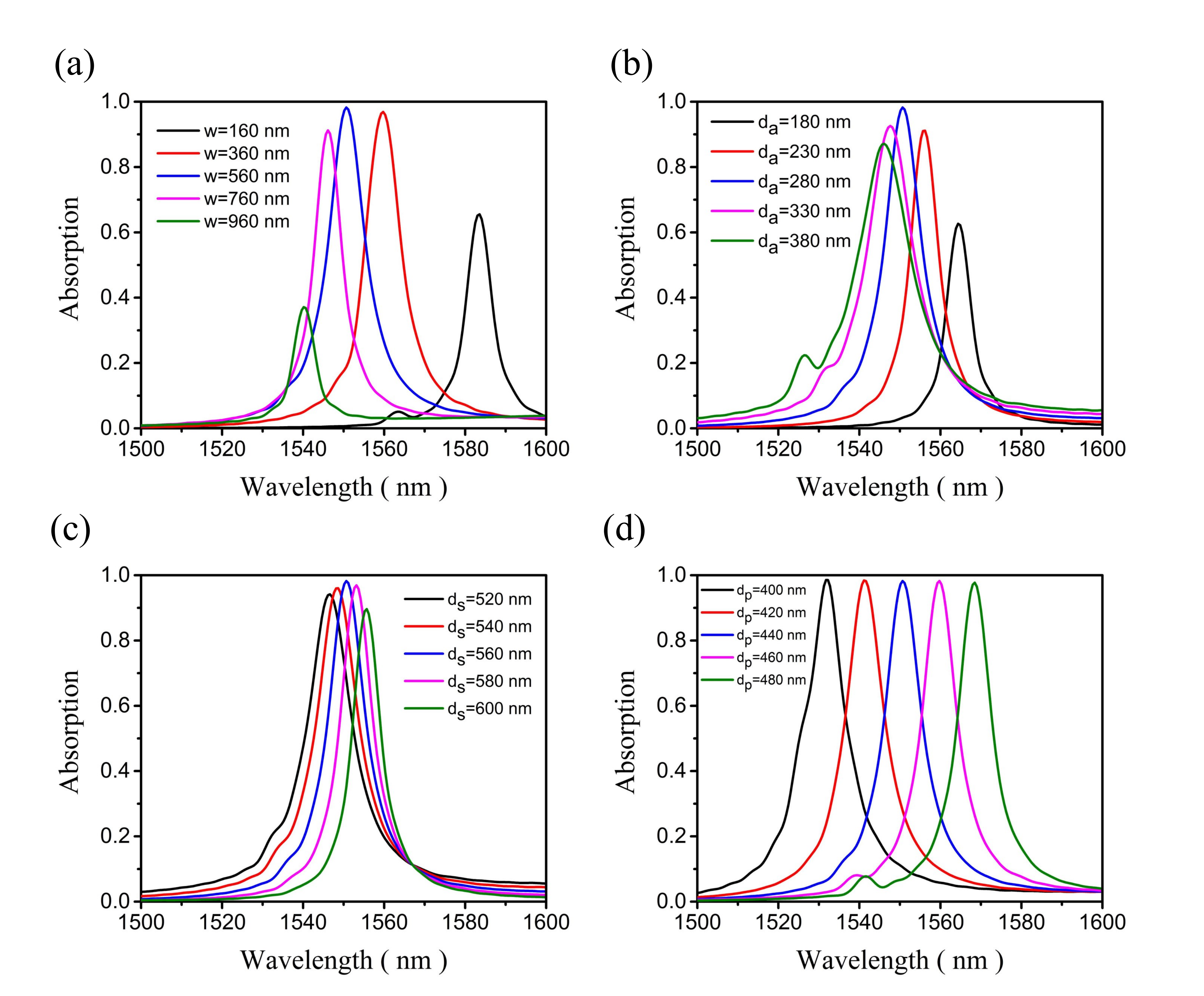}
\caption{The absorption spectra of monolayer graphene with various structural parameters when $P=1250$ nm, $d_1=100$ nm, $d_2=260$ nm and $N=5.5$ pairs. Using different widths (a) and depths (b) of the cross-shaped groove air resonators for $d_s=560$ nm, $d_p=440$ nm,  and using different SiO$_{2}$ layer thickness (c) and PMMA layer thickness (d) for $w=560$ nm and $d_a=280$ nm.}
\end{figure}
Fig.4d shows that the almost perfect absorption of monolayer graphene and the absorption wavelengths are linearly tuned by the thickness of PMMA ($d_p$). The spectra lines are red-shifted from $1532$ nm to $1568$ nm as $d_p$ increases from $400$ nm to $480$ nm. The spectral selectivity of the structure is improved by adjusting the thickness of PMMA ($d_p$), and the feasibility of the experiment is also provided for the designed in the paper\cite{suk2011transfer}. 
In addition, the influence of the thickness of Si ($d_1$) and SiO$_{2}$ ($d_2$) layers in the Bragg mirror are also investigated, as shown in Fig.5a and b, respectively. The peaks of the absorption spectra of monolayer graphene are linearly red-shifted with the increase of the thickness of Si and SiO$_{2}$ due to the change of the gap position, in which the effect of Si are relatively obvious\cite{duguay1986antiresonant}, as can be seen in Fig.5a. Theoretically, the peaks wavelength of the absorption spectra of graphene can be approximately calculated by the equation $\lambda_{0}=2(n_{1}d_{1}+n_{2}d_{2})$, where $n_1$ and $n_2$ are the effective refractive index of Si and SiO$_{2}$ in the Bragg mirror, respectively. From the formula above, we can see that the thickness of Si has a greater influence on the peaks wavelength ($\lambda_0$) than the SiO$_{2}$ layers thickness. But they have minimal effect on the peaks magnitude of the absorption spectra of graphene. 
\begin{figure}[htbp]\label{fig:5}
\centering
\includegraphics[scale=0.32]{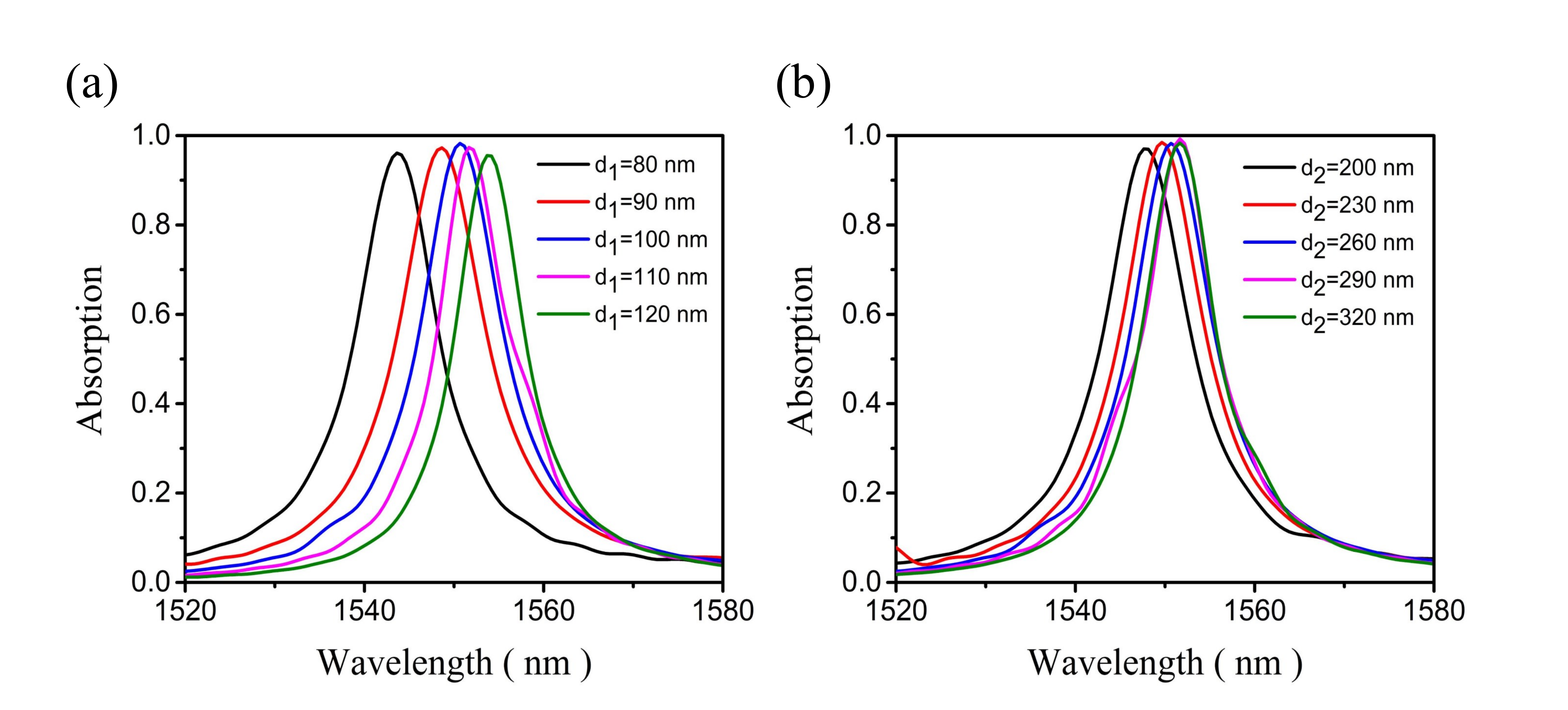}
\caption{Influence of Si layers thickness (a) and SiO$_{2}$ layers thickness (b) in the Bragg mirror on light absorption of monolayer graphene.}
\end{figure}
As depicted in Fig.6, we simulate the absorption of monolyer-graphene (black line) and the whole structure (red line) with increasing the period number ($N$) of the Bragg mirror. It can be found that when the number of period $N>3$, the perfect absorption of the graphene layer can achieve, and then the system almost tends to be stable with the $N$ increases continuously.
\begin{figure}[htbp]\label{fig:6}
\centering
\includegraphics[scale=0.41]{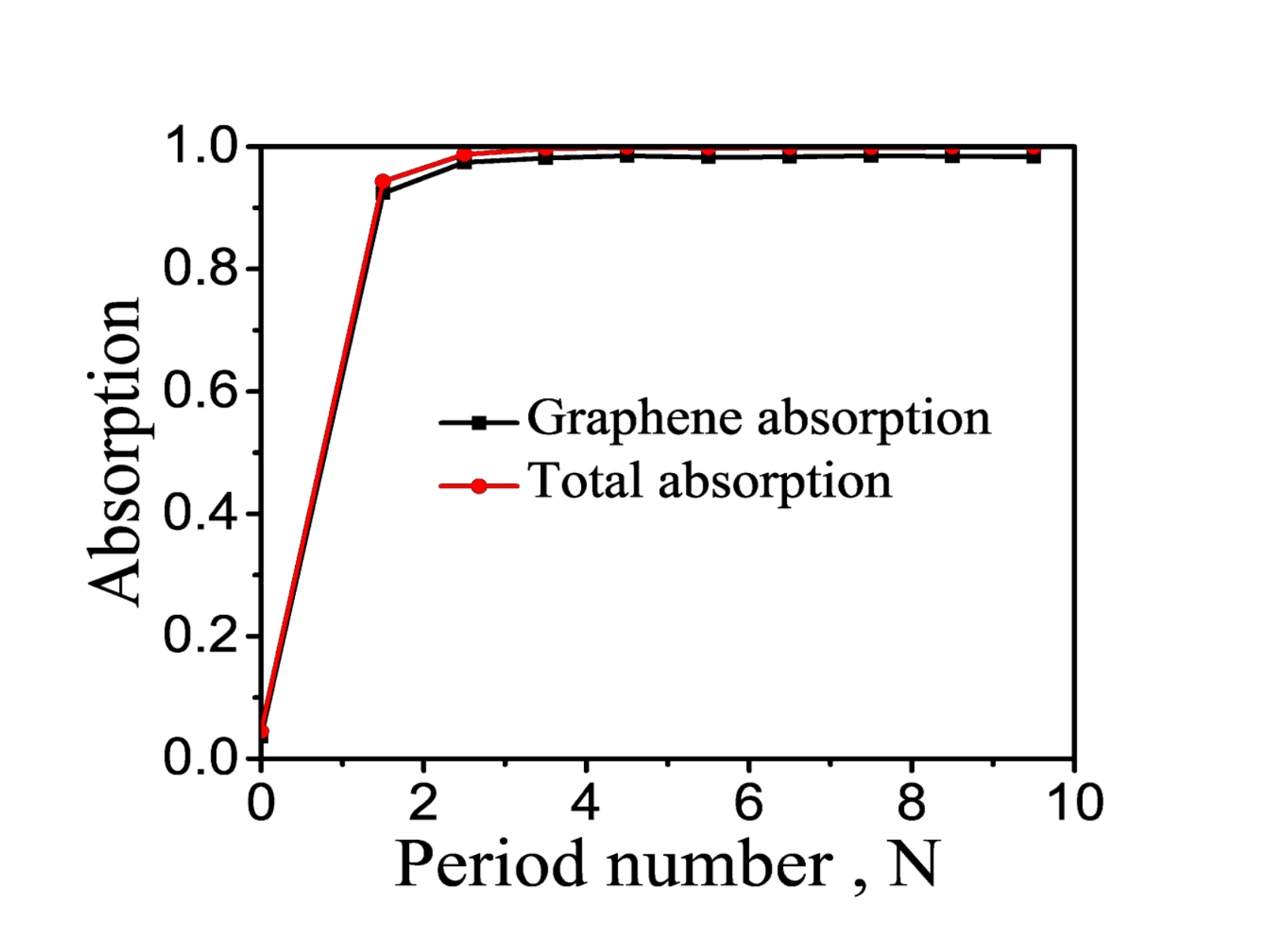}
\caption{Light absorption in the whole structure (red line) and the graphene monolayer (black line) using different period number ($N$) of the Bragg mirror. Other geometric parameters are the same as in Fig.2.}
\end{figure}

Until now, we investigate the various properties of the designed structure at normal incidence. Subsequently, in Fig.7 we show the absorption of the graphene layer as functions of the incident angle and wavelength for TM polarization and TE polarization. As shown in Fig.7a, the wavelengths of the major absorption peaks are almost unchanged with the increase of the incident angle continuously for TM-polarized, mainly due to the insensitivity of the guided mode resonance to the incident angle, it can be valuable in applications on integrated optoelectronic devices. In contrast, as for the TE-polarized, when the incident light is tilted at a certain angle, another resonant mode is stimulated by the incident light, resulting in an additional graphene absorption peak appearing on the absorption spectrum in Fig.7b. Meanwhile, the wavelengths of two graphene absorption peaks are also changing with increasing the tilt angle of incident light. Therefore, we have proved that the designed structure can simultaneously achieve the critical coupling of multiple resonances, which is a major technical index of multispectral optical detection. And these angular characteristics of the structure have potential applications in the field of space optical measurement\cite{qu2016spatially}.
\begin{figure}[htbp]\label{fig:7}
\centering
\includegraphics[scale=0.32]{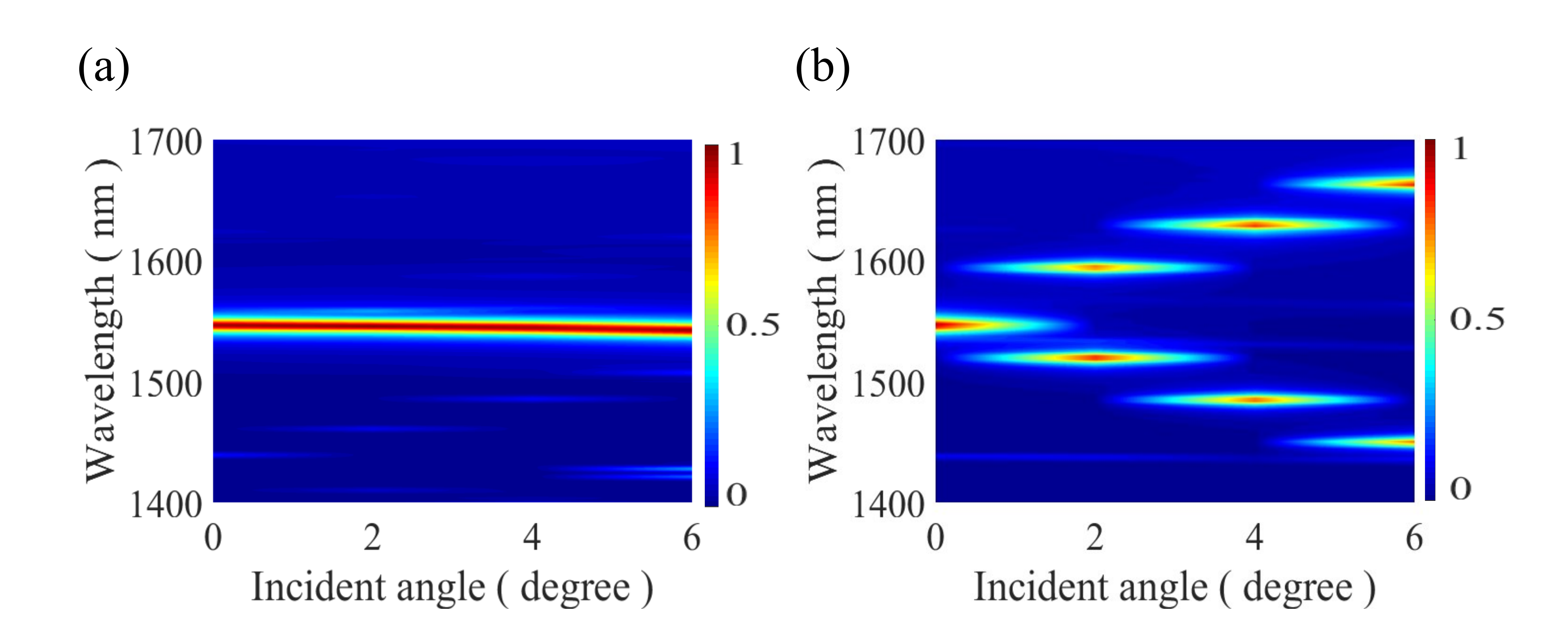}
\caption{Absorption spectra of the monolayer-graphene layer as functions of the wavelengths and incident angle for TM-polarization (a) and TE-polarization (b).}
\end{figure}

\section{Conclusions}
In summary, we investigate a monolayer graphene-based perfect absorption structure with a cross-shaped groove air resonator, in which the absorption of graphene can achieve total absorption (about $99\%$ of the incident light at normal incidence) at telecommunication wavelengths through the mechanism of critical coupling with guided resonance. The modeling work implies that the absorption wavelength of monolayer graphene can be tuned by adjusting the parameters of structure. In other words, the perfect absorption efficiency and spectral selectivity are obtained with attaining critical coupling condition. The results of research work also show that the different polarization modes (TM or TE) have different sensitivity to the incident angle, which lead to their different incident angular tolerance. In addition, the proposed graphene-based perfect absorption structure with a dielectric Bragg mirror together with its design principle can be extended to enhance the absorption of other two-dimensional materials.

\section{Acknowledgements}
The author Xiaoyun Jiang (XYJIANG) expresses her deepest gratitude to her Ph.D. advisor Tao Wang for providing guidance during this project. This work is supported by the National Natural Science Foundation of China (Grant No. 61376055, 61006045 and 61775064), and the Fundamental Research Funds for the Central Universities (HUST: 2016YXMS024).

\end{document}